\documentstyle[twocolumn,aps,prc,epsf]{revtex}
\begin{document}
\draft
\title{ Factorization of Fermion Doubles on the Lattice. }
\author{Pedro J. de A. Bicudo} 
\address{ Departamento de F\'{\i}sica and 
 Centro de F\'{\i}sica das Interac\c c\~oes Fundamentais, 
Edif\'{\i}cio Ci\^encia, \\ 
Instituto Superior T\'ecnico, Av. Rovisco Pais, 
1049-001 Lisboa, Portugal
\\
email: \ fbicudo@alfa.ist.utl.pt 
}
\maketitle
\begin{abstract}  
We address the problem of the fermion species doubling 
on the Lattice.
Our strategy is to factorize the fermion doubles
from the action.
The mass term of the Dirac-Wilson action is changed.
In this case the extra roots which appear in the 
action of free fermions in the moment representation 
are independent of the mass and can be factorized
from the fermion propagator.
However the gauge couplings suffer from the pathological
ghost poles which are common to non-local actions.
This action can be used to find a solution of the 
Ginsparg Wilson relation, which is cured from the
non-local pathology. Finally we compare this factorized
action with solutions of The Ginsparg Wilson relation.
We find that the present is equivalent to the Zenkin 
action, and that is not quite as local as the Neuberger 
action. 
\end{abstract}

\section{introduction}
\par
\begin{figure}
\begin{picture}(240,260)(0,0)
\put(-64,-20){\epsffile{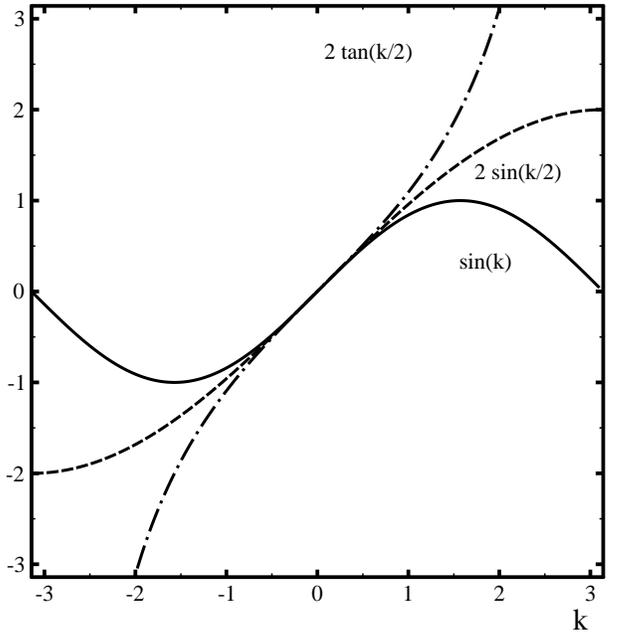}}\end{picture}
\caption{In the Brillouin zone, where k is dimensionless, we show the
functions  $\sin k$, $ 2 \sin {k \over 2} $ and $ 2 \tan {k \over 2} $ .}
\label{Brillouin}
\end{figure}

Fermion doubles 
appear in the lattice whenever one calculates the first derivative with 
the finite difference method and Fourier transforms it. 
This is an old problem \cite{Wilson}, which afflicts lattice 
simulations of any quantum field theory with light Dirac Fermions.
Bosons are healthy
because the second derivative has no doubling problem. This problem can 
be addressed in a simple 1 dimensional lattice. 
In this case, with a lattice spacing $a$, the first derivative of 
a fermion field $\psi$, 
is calculated in the first approximation of the finite difference method,
\begin{eqnarray}
\label{deriva1}
\partial \psi_n &=& {\psi_{n+1}-\psi_{n-1} \over 2a}  
\end{eqnarray}
The Fourier transform of an infinite and periodic lattice is the 
Brillouin zone,
a continuous interval $]-\pi,\pi]$, and the functions are transformed as,
\begin{equation}
\psi_n=\int_{-\pi}^{\pi} {dk \over 2 \pi} e^{i n.k} \ 
\widetilde \psi(k) \ \ , 
\ \ \widetilde \psi(k)=\sum_n e^{-i k.n} \ \psi_n \nonumber \\
\end{equation}
which we substitute in eqs. (\ref{deriva1}) 
and find for the derivative,
\begin{eqnarray}
\partial \psi_n&=& \int_{-\pi}^{\pi} {dk \over 2 \pi} {e^{+ik}-e^{-ik} \over 2 a}
e^{i n.k} \ \widetilde \psi(k)  \nonumber \\
 &=& \int_{-\pi}^{\pi} {dk \over 2 \pi} i{\sin k \over a}
e^{i n.k} \ \widetilde \psi(k)  
\end{eqnarray}
The continuum limit is reached when $a$ vanishes. In this case it is
natural to replace $n \rightarrow x / a$  and $k \rightarrow ap$, and we
get the correct behavior,
\begin{equation}              
{\sin k \over a} \rightarrow p 
\end{equation}
However, one should check all the points where the numerator $\sin k$ 
vanish in the Brillouin zone. In Fig. \ref{Brillouin} we see 
in fact that $\sin k$ has two similar roots in the Brillouin zone, 
at $k=0$ and at $k=\pi$. This is the source of the doubler problem.
\par
In the literature many methods to avoid the doubling of fermion 
species on the lattice are presented.
A simple solution of this lattice problem would be convenient for 
the study of chiral symmetry breaking in QCD and for the study 
of neutrinos in the Standard Model.
Wilson \cite{Wilson} included in the action extra hopping terms 
which projected out the doublers from the onset. However
this was not chiral invariant, and Kogut \cite{Kogut} proposed
the staggered fermions to improve chiral invariance.
Nielsen and Ninomiya \cite{Nielsen} produced a topological no-go 
theorem concerning chiral invariant lattice models with 
local charge conservation and a local action.
To avoid these restrictions, there are some very interesting
solutions in the literature which depart from the simple
Wilson action.
Kaplan \cite{Kaplan} extends the lattice with an extra dimension.
and places the physical fermions in a domain wall.
't Hooft \cite{'t Hooft} recommends to have continuum 
fermions with lattice bosons.
Neuberger \cite{Neuberger} redefines the matrix which is used
in the fermion determinant, and finds a topologically nontrivial 
solution to the Ginsparg-Wilson \cite{Ginsparg} relation. 
This opens a new perspective to the study of chiral symmetry  
\cite{Luscher,Fujikawa}, and the study of 
the chiral anomaly on the lattice.
The method presented in this paper belongs to the class of
non-local actions \cite{Drell,Rebbi,Bietenholz,Zenkin}.
\par
We now show a 1 dimensional example where the doubles are factorized
from the propagator. We find it more intuitive to place
the fermion fields at the vertices of the lattice zones (dual lattice)
because we are tempted to replace the periodic function $\sin k$ by 
the anti-periodic $2 \sin {k \over 2}$ which has a single root.
To get this new function, which is wider in the momentum representation,
see Fig. \ref{Brillouin},
one has to narrow the spacing $2 a$ used in the denominator of the 
derivative. We will later shift our lattice back to the conventional one,
the use of the dual lattice here is not crucial.
The first approximation of the finite difference method is now,
\begin{eqnarray}
\label{derivanova}
\partial \psi_n &=& 
{\psi_{n+{1 \over 2}}-\psi_{n-{1 \over 2}} \over a} \nonumber \\ 
&=& \int_{-\pi}^{\pi} {dk \over 2 \pi} 
i{2 \sin {k \over 2} \over a}
e^{i n.k} \ \widetilde \psi(k)  
\end{eqnarray}
and we obtain the desired function.
We will use the notation $\widehat{\psi}_n$ for the 
interpolated spinor at the site $n$ of the lattice, which
is simply the mean value of the spinors which are 
placed at the vertices of the zone,
\begin{eqnarray}
\label{interpol}
\widehat{\psi}_n
&=& {\psi_{n+{1 \over 2}}+\psi_{n-{1 \over 2}} \over 2} 
\nonumber \\
&=& \int_{-\pi}^{\pi} {dk \over 2 \pi} 
\cos {k \over 2}
e^{i n.k} \ \widetilde \psi(k)  
\end{eqnarray}
We find that the constant term (\ref{interpol}), in sense that it is 
not the first derivative term, has a root at $k=\pi$. This is the 
extra root which was present in the first definition (\ref{deriva1}) 
of the derivative. 

\par
With the interpolated fields (\ref{derivanova}) and (\ref{interpol}), 
we define a free Dirac action in 1 dimension,
\begin{eqnarray}
\label{Dirac}
S&=& a \sum_n \ m \bar{ \widehat \psi}_n \ \widehat \psi_n 
+\bar{ \widehat \psi}_n \gamma \partial \psi_n  \\
 &=& a \int_{-\pi}^{\pi}{dk \over 2\pi}  
{\widetilde{\bar\psi}}(k) \cos{k \over 2} \Bigl( m \cos {k \over 2}
+i \gamma {2 \over a} \sin {k \over 2} \Bigr) \widetilde\psi(k) \nonumber 
\end{eqnarray}
were the $\delta$ relations,
\begin{eqnarray}
\label{deltas}
\int_{-\pi}^{\pi} {dk \over 2 \pi} e^{i k.(n-n')} &=&\delta_{n,n'} 
\nonumber \\ 
\sum_n e^{-i n.(k-k')} &=& \sum_m 2 \pi \delta(k-k'+2 \pi m)
\end{eqnarray}
were used. 
We can check that both the mass term (one double root)
and the derivative term (two different roots) of the Dirac action 
(\ref{Dirac}) have two roots. Thus this action complies with the
Nielsen and Ninomiya topological theorem \cite{Nielsen}.
We see that there are still 2 roots , 
${2 \over a}\tan{k \over 2}=m$ and $k=\pi$ 
in the Lagrangian (\ref{Dirac}), but only the first root
is dynamical in the sense that it depends on the fermion mass. 
Nevertheless it is possible to define a propagator $G$ which only shows 
the correct pole in the denominator, and no double,
\begin{eqnarray}
G_{nn'} &\equiv&  
\langle \widehat \psi_n \bar{\widehat \psi}_{n'}\rangle  \nonumber \\ 
 &=&  \int_{-\pi}^{\pi}{dk \over 2\pi a}  \ { e^{ik.(n-n')}  \over
  m + i \gamma {2 \over a} \tan {k \over 2}  }
\label{propagator1}
\end{eqnarray}  
where the $\cos^2{k \over 2}$ in the numerator, which cancels the extra 
root in the denominator, comes from the interpolation of the fermion 
fields in the definition of the propagator.
The extra root is replaced by a pole.
The propagator (\ref{propagator1}) has the right continuum limit,
\begin{equation}              
i \gamma {2 \over a}\tan{k \over 2} +m \rightarrow i \not p + m . 
\end{equation}
For a finite spacing $a$, we can see in Fig. \ref{Brillouin} that this 
propagator also has the peculiar property $2 \tan{ k \over 2} > k$,
which increases the effect of the ultraviolet cutoff ${\pi \over a}$ 
in fermions. This is in contradistinction with $\sin k < k$. 
\par
In the next Section II we define the factorization procedure
in detail, for a Lattice in $d$ dimensions. In section III
gauge invariance is implemented. In section IV the
factorization is adapted to the Ginsparg-Wilson relation.
The conclusion is in Section V. In Appendix A an
equivalent factorization is studied, and in Appendix B
we study lattices where fermions and antifermions are 
in separate sites.

\section{factorization of the doubles in $d$ dimensions}
\par
We now extend the action (\ref{Dirac}) to the $d$ dimensional case,
and factorize the doubles from the propagator and the fermion
determinant. In this section we study free fermions which can be
exactly solved in the momentum representation. However the factorizing
procedure must be applicable in position space, in order to include
eventually the gauge fields, and to compute the propagator
and the determinant with the standard lattice techniques.
At this point we have to leave the dual lattice which 
was used just for the introduction. 
In order to write the interpolated spinors, it is convenient to 
define the 
class of vectors $\vec \mu$, with components $\mu_l=+1\, , \, -1$.
In $d$ dimensions, the interpolated spinors are now extended to,
\begin{eqnarray}
\displaystyle
\label{interpoln}
\widehat{\psi}_{\vec n}
&=& 
{1 \over 2^d} \sum_{\mu_l} \psi_{ \vec{n}+{1 \over 2} \vec{\mu} }
\ , \ \
\partial_j\psi_{\vec n} =
{1 \over a \ 2^{d-1}} 
\sum_{\mu_l} \mu_j \psi_{ \vec{n}+{1 \over 2} \vec{\mu} }
\end{eqnarray}
To translate the fields back to the standard positions $\vec n$ 
we apply a shift of $-\vec \mu' / 2$ on the action (\ref{Dirac}).
This shift leaves the momentum representation of the action invariant. 
The action is now,
\FL
\begin{eqnarray}
\label{naction2}
S &=& \left({a \over 4}\right)^d \sum_{\vec n,\vec \mu,\vec \mu'} 
\bar{ \psi}_{\vec n} \Bigl( m+
{2 \over a }  \sum_j \, \mu'_j \, \gamma_j  \Bigr)
\psi_{\vec n + {\vec \mu' - \vec \mu  \over 2}} 
\end{eqnarray}
where we use the convention of hermitean Dirac matrices $\gamma_i$.
The action (\ref{naction2}) can be written in a matricial form,
\begin{eqnarray}
\label{matrixaction}
S=a^d \sum_{\vec n,\vec n'} m \bar \psi_{\vec n} I_{\vec n,\vec n'} 
\psi_{\vec n'}+ a^d \sum_{j,\vec n,\vec n'}
\bar \psi_{\vec n}\gamma_j {\partial_j}_{\vec n,\vec n'}\psi_{\vec n'}
\end{eqnarray}
where the sites are labeled by the $\vec n, \vec n'$, which
are $d$-dimensional vectors with integer components.
Like the Wilson fermion action \cite{Wilson}, this action
includes extra hopping terms. In the present case the 
chiral non-invariant terms are all included in the mass term.
In eq. (\ref{matrixaction}) the matrices
$I$ and $D_j$ are translational invariant, 
in the sense that the matrix elements only
depend on the distance to the diagonal. The Fourier transform
of these matrices depends on a single momentum,
\begin{equation}
I_{\vec n,\vec n'}=\int_{-\pi}^{\pi} {d^dk \over (2 \pi)^d}
e^{i (\vec n'-\vec n).\vec k} \ \widetilde {I}(\vec k) \  ,
\end{equation}
and from eq. (\ref{interpoln}) we get the Fourier transforms,
\begin{eqnarray}
\widetilde {\partial_j}(\vec k)&=& i  {2 \over a} \tan {k_j \over 2}
\widetilde I(\vec k) \ , \ \
\widetilde I(\vec k)=\prod_l \cos^2 {k_l \over 2} \ .
\end{eqnarray}
We define the
quark propagator $G$ with the Green equation,
\begin{eqnarray}
\label{Green}
\Bigl( mI+\sum_j \gamma_j \partial_j \Bigr)  G=I \ . 
\end{eqnarray}
This Green equation (\ref{Green}) is an extended
version of the normal one. Here we replace the
usual identity matrix in the right hand side by
the matrix $I$. The action with no doubles 
is finally defined,
\begin{eqnarray}
\label{finalaction}
S= a^d &&\sum_{\vec n,\vec n'} \bar \psi_{\vec n} 
D_{\vec n,\vec n'} \psi_{\vec n'} \ ,
\nonumber \\
D&&=G^{-1}=m+\sum_j\gamma_j \, I^{-1}\partial_j
\end{eqnarray}
where the matrix $M$, in the limit of massless quarks is
chiral invariant.
Like the Neuberger matrix \cite{Neuberger},
the final matrix $D$ is redefined from the
initial action (\ref{naction2}).
Using the delta relations (\ref{deltas}) 
we find the momentum representation of the action,
\begin{eqnarray}
\label{final momentum}
S=a^d&&\int_{-\pi}^{\pi} {d^dk \over (2\pi)^{d}}
\widetilde{\bar \psi}(k) \widetilde D(k)
\widetilde \psi(k) \ ,
\nonumber \\
\widetilde D(\vec k)&&=\widetilde I^{-1}(\vec k)
\Biggl[ m \widetilde I(\vec k) +\sum_j \gamma_j 
\widetilde {\partial_j}(\vec k)
\Biggr] 
\nonumber \\ &&=
 m+i  {2 \over a} \sum_j \gamma_j 
\tan {k_j \over 2} \ .
\end{eqnarray}
We see that the new action has no extra roots 
and has the correct continuum limit. This result can be
compared with the literature. Except for the mass term,
we recover the Zenkin action \cite{Zenkin}, which was 
derived in the formalism of a path integral with 
the Weyl quantization.
\par
\section{Coupling to the gauge fields}
\par
We now include the coupling of the fermions to the gauge fields, using
the Wilson \cite{Wilson} prescription. 
In order to have a gauge invariant action, the fermion
operators $\bar \psi$ and $\psi$ are usually linked by gauge 
transformation paths when they are in different sites. In the case 
of a elementary link $\vec j$ , 
\begin{equation}
U_{\vec n ,\vec n +\vec j}= 
e^{i \vec B(\vec n + \vec j / 2) . \vec j} \ , 
\  \vec B ={1 \over 2} a \, g  \sum_b \lambda_b . 
 \vec A_b
\end{equation}
where $ \lambda_b $ are the generators of the gauge group, 
and the gauge-invariant
version of $\bar \psi_{\vec n } \psi_{\vec n + \vec j}$ is,
\begin{equation}
\bar \psi_{\vec n }e^{i \vec B . \vec j} \psi_{\vec n + \vec j}.
\end{equation}
In the present case we have to connect $\bar \psi_{\vec n}$ and 
$\psi_{\vec n + {\vec \mu - \vec \mu ' \over 2}}$
of action (\ref{naction2}) by a gauge link. Because each of
the $d$ components of $(\vec \mu - \vec \mu ' ) / 2$
can be equal to $\pm 1$ or $0$, this gauge link can be constituted
by a product $\Pi \, U$ of up to $d$ concatenated elementary orthogonal 
elementary links.
Then there are different paths which connect
$\bar \psi_{\vec n}$ and $\psi_{\vec n'}$, and it is convenient
to use the average of all the possible paths, or to choose 
a single path in a random way. With the links included, the $I$ and 
$\partial$ 
terms in eq. (\ref{matrixaction}) are gauge invariant an so is the new 
action (\ref{finalaction}).

\par
Including the gauge coupling, the fermion part of the action
(\ref{finalaction}) is extended to,
\begin{eqnarray}
\label{gaugeaction}
S&=a^d &\sum_{\vec n,\vec n'} m \bar \psi_{\vec n} D _{\vec n,\vec n'} 
e^{i  \vec B_{n,n'} \cdot ( \vec n'- \vec n )} \psi_{\vec n'}  \ .
\end{eqnarray}
In the continuum limit $\vec B$ tends to a vanishing constant. 
This action (\ref{gaugeaction})
is the one that we propose for the study of light fermions.
\par
In the limit of a constant $\vec B$ and using the $\delta$ relations 
(\ref{deltas}) the action (\ref{gaugeaction})
can be diagonalized in the momentum representation,
\begin{equation}
\label{minimal}
S=a^d \int_{-\pi}^{\pi} {d^dk \over (2 \pi)^d}
\widetilde{\bar \psi}(k) 
\widetilde {D}(\vec k +\vec B) 
\widetilde{ \psi}(k) 
\end{equation}
which is the minimal coupling of the gauge fields to the fermions. 
\par
Computing either the propagator or the fermion determinant,
is crucial to directly measure chiral symmetry breaking.
In the lattice, the Green equation for the propagator,
\begin{eqnarray}
\label{Greengauge}
&\sum_{\vec n''}
\Bigl( mI_{\vec n,\vec n''}+\sum_j \gamma_j
{\partial_j}_{\vec n,\vec n''}\Bigr)
(\Pi \,  U )_{\vec n,\vec n''} 
G_{\vec n'',\vec n'}&
\nonumber \\
&=I_{\vec n,\vec n'} (\Pi \,  U )_{\vec n,\vec n'}
 \ . &
\end{eqnarray}
is perfectly defined and can be solved numerically
with the standard lattice methods \cite{Kogut2}. Although
$I$ is not the identity matrix, its nonvanishing
matrix elements are very close to the diagonal. The
fermion determinant $\left|D \, U\right|$ is also calculable.
As for the pure gauge term, the Wilson action is
unchanged in this case.
\par
However the pole in the action, which was introduced
to solve the doubler problem, raises a new difficulty 
\cite{Pelissetto} when the gauge coupling is included.
This can be understood when we extract pertubatively 
the coupling of the fermion to a gauge field from eq. 
(\ref{minimal}). The same result can be extracted 
from the vector Ward identity, and we get,
\begin{eqnarray}
\Gamma_j &=& {\partial \over \partial p_j} \widetilde D( \vec k)
\nonumber \\
&=& i \, \gamma_j \left( 1 + \tan^2 {k_j \over 2} \right).
\end{eqnarray}
Now the pole in the fermion Lagrangian (\ref{final momentum}) will 
introduce a pole in the coupling say of a photon to an electron.
Hence the doublers, which were expelled from the free theory,
reappear \cite{Pelissetto} through the gauge independent interaction.
This problem is common to the  non-local actions, except for the 
Bietenholz-Wiese perfect action \cite{Bietenholz}.

\section{A solution to the Ginsparg-Wilson relation}
\par
We can address at the same token \cite{Chiu} the coupling 
problem \cite{Pelissetto}, and the Ginsparg-Wilson 
\cite{Ginsparg} relation,
\begin{equation}
D \gamma_5 + \gamma_5 D = 2 r D \gamma_5 D ,
\end{equation}
where $r$ is a constant of the order of the lattice spacing $a$.
In the case where the action complies with the
Ginsparg-Wilson relation,
L\"uscher \cite{Luscher} showed that the lattice has
redefined a chiral symmetry, which includes the chiral anomaly. 
Moreover, in the continuum limit of $a \rightarrow 0$, the chiral 
symmetry and chiral anomaly of conventional field theory
could be recovered \cite{Fujikawa}.
Here we will follow the Chiu-Zenkin \cite{Chiu} prescription to 
solve the Ginsparg-Wilson relation, which has a linear version 
for the inverse Dirac matrix,
\begin{equation}
\label{inverse}
D^{-1} \gamma_5 + \gamma_5 D^{-1} = 2 r \gamma_5  \ .
\end{equation}
A solution of this equation is the sum of
an homogeneous solution plus the particular one. 
Any chiral invariant matrix $D^{-1}_c$ is a solution of the 
homogeneous equation. We will use the solution
of eq. (\ref{Greengauge}) with a vanishing fermion mass $m$. 
A simple particular solution is $ D^{-1}_p =r$. Our solution is then,
\begin{eqnarray}
\label{ginsparg solution}
D^{-1} &=& D_c^{-1}+D_p^{-1}  \ \ ,
\nonumber \\
D&=&{ D_c \over 1 + r  D_c}
\nonumber \\
 &=& {1\over I + r \sum_j \gamma_j \partial_j}
     \sum_j \gamma_j \partial_j     \ \ .
\end{eqnarray}
In this matrix $D$, the poles are factorized, this is a simple
extension of eq.(\ref{Greengauge}). If we turn off the gauge fields,
the Fourier transform can be computed,
\begin{eqnarray}
\label{ginsparg momentum}
\widetilde D&=&{ i {2 \over a} \sum_j \gamma_j  \tan {k_j \over 2}
 \over 1 + r \ i {2 \over a} \sum_j \gamma_j  \tan {k_j \over 2}}  
\nonumber \\
&=&{\sum_j \ i \, {2 \over a}  \gamma_j  \tan {k_j \over 2}
+  r {4 \over a^2}  \tan^2 {k_j \over 2}
\over 1 + r^2   {4 \over a^2} \sum_j  \tan^2 {k_j \over 2}}\ \ ,
\end{eqnarray}
which now has a single root, with no other doubles, and no 
pole in euclidian space. 
\begin{figure}
\begin{picture}(220,380)(0,0)
\put(5,40){
\put(10,-20){\epsffile{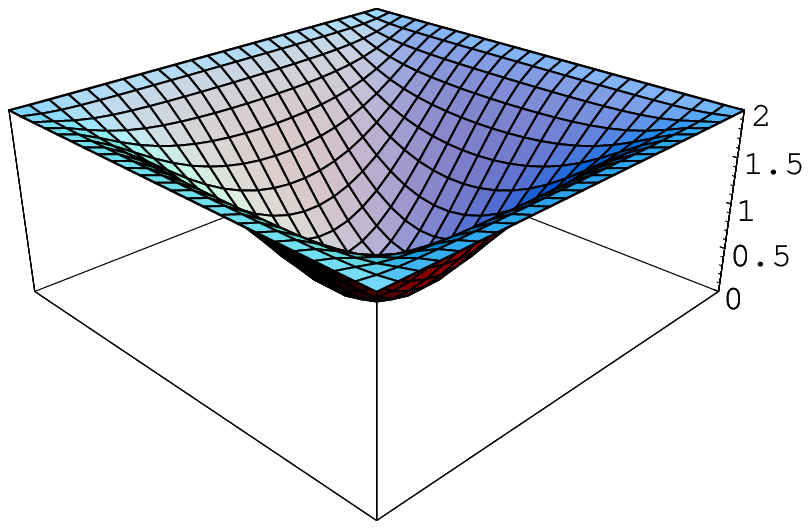}}
\begin{picture}(240,160)(0,0)
\put(0,0){(a)}
\put(40,00){$k_1$}
\put(195,00){$k_2$}
\put(0,40){$-\pi$}
\put(90,-20){$\pi$}
\put(130,-20){$-\pi$}
\put(215,40){$\pi$}
\end{picture}
}
\put(5,220){
\put(10,-20){\epsffile{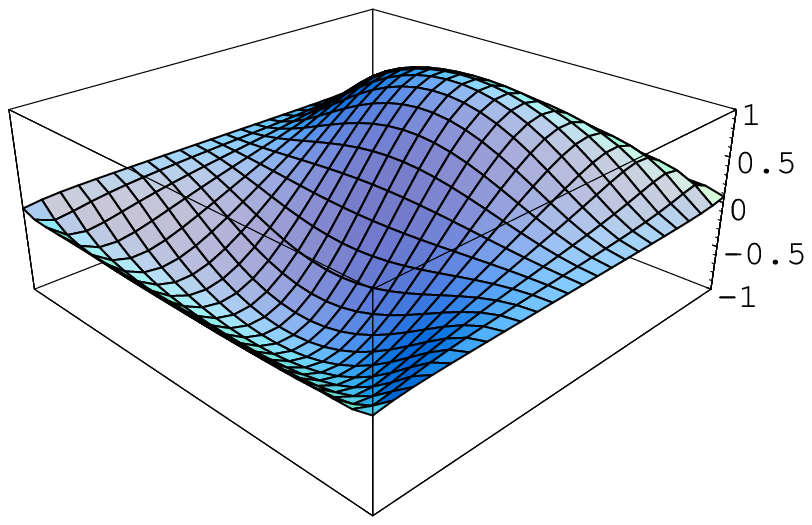}}
\begin{picture}(240,160)(0,0)
\put(0,0){(b)}
\put(40,00){$k_1$}
\put(185,00){$k_2$}
\put(0,40){$-\pi$}
\put(90,-20){$\pi$}
\put(127,-20){$-\pi$}
\put(205,37){$\pi$}
\end{picture}
}
\end{picture}
\caption{
On a $24$x$24$ Brillouin lattice with parameters $a=r=1$ 
we respectively show the Dirac scalar 
component (a), and the $i\gamma_2$ component (b) of 
the Fourier transform of the free action.
}
\label{3D plot}
\end{figure}
It is simpler when $ r=a/2 $ and in the case where all 
components $k_j$ vanish except for $k_i$.
In this case eq. (\ref{ginsparg momentum}) is 
equal to $\left(i {1 \over a} \gamma_i  \sin k_i  
+ { 1 \over r }\sin^2 {k_i \over 2} \right)$,
which matches the Wilson action. See Fig. \ref{3D plot}
for a plot of the free action in momentum space.
\par
In the case where the fermion has a mass $m$ from the onset,
a natural choice for the action consists in using the Dirac matrix
$D$ of eq. (\ref{ginsparg solution}) where we replace the homogeneous
solution by the full solution of eq. (\ref{Greengauge}), including
the mass term. The action is then, 
\begin{equation}
D = {1\over I + r \ \sum_j \gamma_j \partial_j + r \ I m }
    \left(\sum_j \gamma_j \partial_j +I m \right) \ \,
\end{equation}
and in the free fermion case the Fourier transform is,
\begin{eqnarray}
\widetilde D&=&{ i  {2 \over a}  \sum_j \gamma_j  \tan {k_j \over 2}
+  r {4 \over a^2} \sum_j \tan^2 {k_j \over 2} +m+r \, m^2
\over 1 + 
r^2   {4 \over a^2} \sum_j  \tan^2 {k_j \over 2}+ r^2 m^2} \ .
\end{eqnarray}
\section{Comparing with the Neuberger Solution}
\par
We now return to the relevant chiral limit of $m=0$, and compare
the factorized action with the literature. 
The present action is formally very simple, and it is equivalent 
to the Zenkin action. It is interesting to compare it with 
the Neuberger action which is presently the most 
promising solution of the Ginsparg-Wilson relation. 
\par
The solution (\ref{ginsparg solution}) has two interesting regions
in the Brillouin zone. In the center we have the single root.
There the chiral breaking term vanishes and 
$D \simeq i \not \hspace{-.1cm} p$. 
The other interesting region is the whole boundary of the Brillouin 
zone, where the chiral invariant term vanishes 
and $D \simeq 1/r$. Thus we have a whole surface
where $1-rD=0$ and this implies that the Chiu-Zenkin
condition \cite{Chiu}, $det(1-rD)=0$ is verified. This
is a necessary condition \cite{Chiu} for the topological
\cite{Luscher} index
\begin{equation}
tr{\gamma_5 D}=$index$ \ D = n_--n_+
\end{equation} 
to be nonvanishing.
The Neuberger \cite{Neuberger}
action  $D_{\cal N}$
has the same single root, and has $2^d-1$ points
\cite{Fujikawa} (at the traditional position of the doublers) 
where $D_{\cal N}= 1/r$.
\par
\begin{figure}
\begin{picture}(220,180)(0,0)
\put(0,50){
\put(10,-20){\epsffile{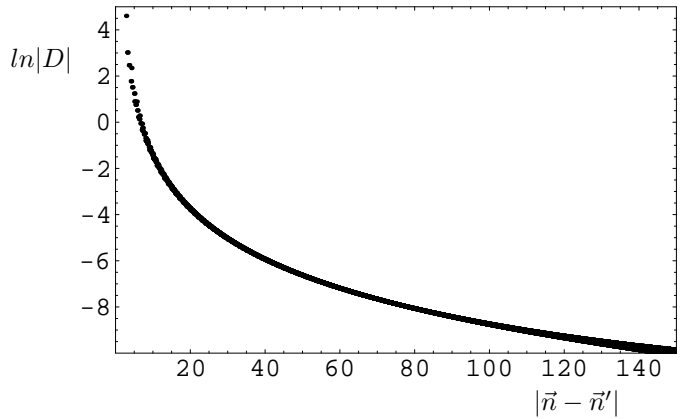}}
\begin{picture}(240,160)(0,0)
\put(-13,100){$ln|D|$}
\put(185,-30){$|\vec n-\vec n'|$}
\end{picture}
}
\end{picture}
\caption{
On a $500$x$500$ lattice with parameters $a=r=1$ and in the case
of free fermions we show $ln(|D|_{\vec n-\vec n'})$ as a function 
of $\vec n-\vec n'$.
}
\label{logarithmic}
\end{figure}

It is also straightforward to study the locality of the 
action $D$. In the literature we can find different 
definitions of a local action. Nielsen and Ninomiya called
local an action with continuous first derivative
in momentum space. The present action $D$ falls in this class. 
L\"uscher is more restrictive \cite{local}, and calls local an 
action of class $C^\infty$, in this class the interactions
in position decrease at least exponentially with distance. 
The Neuberger action belongs to this class. The present 
action does not, except for dimension $d=1$, because at the 
usual position of the doubles its second derivatives do not exist. 
In $d=2$ dimensions  
we Fourier transform back to the position
space the free action $\widetilde D(k)$, and draw
\begin{equation}
|D|_{\vec n,\vec n'}=|D_0+i\sum_j \gamma_j D_j|_{\vec n,\vec n'}
=\sqrt{\sum_{j=0}^d\left|{D_j}_{\vec n,\vec n'}\right|^2} 
\end{equation}
in a logarithmic plot as a function of $|\vec n - \vec n'|$,
see Fig. \ref{logarithmic}. We find that in the limit of 
large distances $|D|$ is approximately equal to
$exp(-2 |\vec n - \vec n'|^{0.3})$, so the interactions decrease faster
than a rational function. 
The finite range interactions, like the Wilson action, are 
usually called ultralocal \cite{ultralocal}, but there are
no solutions of the Gisnparg-Wilson relation with an ultralocal
$rD$ . 
  
\par
The index of our $D$ has been recently computed by Chiu
\cite{Chiu,private}, who found that $D$ is topologically
trivial. The action of Neuberger $D_{\cal N}$ complies
with the index relation. Thus it seems that the action
$D_{\cal N}$ is the best candidate to simulate continuum QCD,
including the axial anomaly. Because $D$ has  apparently no 
chiral anomaly, it would be interesting to apply it to the study 
of chiral symmetry breaking on the lattice with a single flavor. 
\par
In continuum QCD the Attiyah-Singer index theorem is
verified, but in the lattice this is still an open 
problem. According to Niedermayer \cite{Niedermayer}, 
this is scale dependent.
If the topological object is larger than the scale of the
action then the index theorem should be complied with, however
if the topological object has a smaller scale, the index 
relation is not verified. In Fig.  \ref{logarithmic} we see
that a gauge configuration with finite topological should
extend itself for more than 20 or 40 lattice spacings,
in order to be larger than the scale of the interaction.
In the case of $D_{\cal N}$, the scale of the fermionic
matrix is just a few lattice spacings, an it was checked
that $D_{\cal N}$ complies with the index relation even
in a small 6x6 lattice \cite{Chiu check}.
So it would be interesting to use very large gauge 
topological objects to check if they would produce a 
finite index for the fermionic matrix $D$. 
\section{Conclusion}
\par
With the aim to simulate euclidian QCD, we find a
lattice action where the extra roots corresponding 
to doubled fermions are factorized from 
the fermion propagator and from the fermion determinant.
Like the Wilson fermion action \cite{Wilson},
extra hopping terms are included in a initial action.
The gauge invariant action with a conventional chiral 
symmetry has undesired poles in the coupling of a gauge 
boson to the fermion, 
but the action with the Ginsparg-Wilson-L\"uscher 
\cite{Ginsparg,Luscher}
symmetry does not suffer from the same pathology. 
The factorized fermionic action $D$ turns out to be 
equivalent to the Zenkin \cite{Zenkin,Chiu} action.
This action complies with the Ginsparg-Wilson
\cite{Ginsparg} relation and with the Chiu-Zenkin 
\cite{Chiu} criterium. However it is only weakly 
local and it has been checked that it is 
topologically trivial, at least in small lattices.
The study of this action might help to understand
the index relation on the lattice.
We conclude that the interest in the present action 
is just theoretical because the Neuberger action 
\cite{Neuberger} is a better candidate to the  
simulation of QCD on the lattice. 

\acknowledgements
\par
I thank Michael Creutz, Jean-Fran\c cois Lagae,
Jeffrey Mandula and Hugh Shananan for introducing me to
the lattice fermion problem during the 1996 Como II 
conference, and Jo\~ao Seixas and Victor Vieira from CFIF
for discussions on lattice techniques. I am indebted 
to Michael Creutz and to Micha Polikarpov for their 
remarks and suggestions. This paper was improved
by the comment of Wolfgang Bietenholz who pointed out 
the problems and solutions of non-local lattice actions, 
and by the comment of Ting-Wai Chiu who called my attention
the action of Sergei Zenkin.
\appendix
\section{Other possible solutions with different lattices}
The factorization of doubles can also be applied to different 
lattices, where the fermion and antifermion fields are placed in 
different sites.
In these cases, unlike the solution in Section II,
interpolations are unavoidable in the action, and
we no longer have to comply with the Nielsen and
Ninomiya theorem. It is possible to find actions without
doubles and without the gauge coupling pathology.
The gauge links are chosen to
directly connect the antifermion sites with the nearest neighbor 
fermion sites, it is not
necessary to concatenate a few links in order to reach a
fermion from an antifermion.
\par
An example of a lattice where the same factorization procedure
works is a alternate lattice where the $\psi$ and $\bar \psi$ fields 
are placed at alternate sites. 
The link structure is simple, however the gauge lattice is $2^d$ 
bigger than the lattice of section III.
\par
We will study in more detail another example, where the anti fermion
fields $\bar \psi_{\vec n}$ are placed in the normal lattice and the
fermion fields $\psi_{\vec n+ \vec \mu/2}$ are placed in the dual
lattice (or vice versa).
For $d>1$, the lattice is composed by plaquettes with $4$ 
elementary links and the Wilson pure gauge action can be used here.
However the number of links per zone, the number of links which meet 
in a vertex and the link length differ from the case of Section III. 
In the case of Section III these numbers were respectively $d$,  $2d$,
and $a$ , while in the present case these numbers are respectively 
$2^d$,  $2^d$ and $a \sqrt{d}/2$. In the case of $d=4$ the number
of links is now increased by a factor of $4$ and the number of links
which meet at a vertex is now doubled.
Applying the factorization procedure to the free fermion action,
\FL
\begin{eqnarray}
\label{naction3}
S &=& \left({a \over 2}\right)^d \sum_{\vec n,\vec \mu } 
\bar{ \psi}_{\vec n} \Bigl( m+
i{2 \over a }  \sum_j \, \mu_j \,\gamma_j  \Bigr)
\psi_{\vec n + {\vec \mu \over 2}} 
\\ \nonumber
&&
\widetilde {\partial_j}(\vec k)= i  {2 \over a} \tan {k_j \over 2}
\widetilde I(\vec k) \ , \ \
\widetilde I(\vec k)=\prod_l \cos {k_l \over 2} \ ,
\end{eqnarray}
we find the same momentum representation 
(\ref{ginsparg momentum}) in the final action.
\par

\end{document}